\documentclass[aps,twocolumn,groupedaddress,nofootinbib,amsmath,amssymb]{revtex4}
\usepackage{dsfont}
\usepackage{bm}
\begin{document}

\title{Rotating AdS black hole stealth solution in $D=3$}

\author{Mokhtar Hassa\"{\i}ne}\email{hassaine-at-inst-mat.utalca.cl}
\affiliation{Centro de Estudios Cient\'{\i}ficos (CECs), Valdivia, Chile.\\
Instituto de Matem\'atica y F\'{\i}sica, Universidad de
Talca, Casilla 747, Talca, Chile.}

\begin{abstract}
We show that the rotating asymptotically anti de Sitter black hole solution of new massive gravity
in three dimensions can support a static stealth configuration given by a conformally coupled scalar field. By static stealth configuration, we
mean a nontrivial time independent scalar field whose energy-momentum tensor vanishes identically on the rotating black hole
metric solution of new massive gravity. The existence of this configuration is rendered possible because of the presence of a gravitational
hair in the black hole metric that prevents the scalar field to be trivial. In the extremal case, the stealth scalar field diverges at the horizon as it occurs for the
conformal scalar field of the Bocharova-Bronnikov-Melnikov-Bekenstein solution in four dimensions.
\end{abstract}

\maketitle
%%%%%%%%%%%%%%%%%%%%%%%%
\section{Introduction}
%%%%%%%%%%%%%%%%%%%%%%%%
The fundamental tenet of general relativity is the manifestation of
the curvature of spacetime produced by the presence of matter. This
phenomena is encoded through the Einstein equations that relate the
Einstein tensor or any other gravity tensor ${\cal G}_{\mu\nu}$ to the
energy-momentum tensor $T_{\mu\nu}$ that arises form the variation of the matter with respect to the metric,
\begin{eqnarray}
{\cal G}_{\mu\nu}=8\pi G\,T_{\mu\nu}.
\label{Einsteineqs}
\end{eqnarray}
Since the energy-momentum tensor depends explicitly on the metric,
both sides of the equations must be solved simultaneously. However,
one can ask if, for a fixed geometry solving the
vacuum gravity equations, is it possible to find a matter source
coupled to this spacetime that does not affect the geometry. Concretely, this
problem consists in examining a particular solution of the Einstein
equations (\ref{Einsteineqs}) where both sides of the equations
vanish, i. e.
\begin{eqnarray}
{\cal G}_{\mu\nu}=0=8\pi G\,T_{\mu\nu}.
\label{StealthEinsteineqs}
\end{eqnarray}
In three dimensions, such gravitationally undetectable solutions called {\it stealth configurations}
have been obtained in \cite{AyonBeato:2004ig} for a nonminimally coupled scalar field with the
BTZ (Banados-Teitelboim-Zanelli) metric \cite{Banados:1992wn}. That is, the gravity action is described by the standard
Einstein-Hilbert  action with a
negative cosmological constant  while the matter source is
given by
\begin{eqnarray}
S_M=-\int d^{3}x\sqrt{-g}\left[\frac{1}{2}
\partial_{\mu}\Phi\partial^{\mu}\Phi+\frac{\xi}{2}R\Phi^2+U(\Phi)\right],
\label{sfnc}
\end{eqnarray}
where $\xi$ denotes the nonminimal coupling parameter, $R$ the scalar curvature
and $U(\Phi)$ is a potential term. As shown in \cite{AyonBeato:2004ig}, a stealth configuration on the
BTZ metric can be obtained for any value of $\xi$ provided the scalar field being nonstatic (time dependent)
and the angular momentum of the BTZ metric being switched off. The same problem has also been considered in higher dimensions
for the same stealth matter source but with a flat geometry \cite{AyonBeato:2005tu}. Recently, black hole stealth configurations
have been obtained with a nonminimal scalar field with a mass term, $U(\Phi)\propto \Phi^2$, and  where the gravity side is described
by the  Einstein-Gauss-Bonnet gravity \cite{Gaete:2013ixa} or its Lovelock
generalization \cite{Gaete:2013oda}.

Here, we will consider the problem of three-dimensional stealth configuration for which the
gravity side of the stealth equations (\ref{StealthEinsteineqs}) is given by the so-called new massive gravity
\cite{Bergshoeff:2009hq}. This alternative three-dimensional gravity theory has raised a lot of interest in the last five years due to interesting properties. Indeed, this theory presents the advantage
of being at the linearized level equivalent to the unitary
Fierz-Pauli theory for free massive 2 spin gravitons and hence it is
a good candidate for a quantum theory of gravity. Up to now, there only exist
two black hole solutions for the new massive gravity equations: an asymptotically AdS spacetime \cite{Oliva:2009ip} and a solution with a Lifshitz dynamical exponent
$z=3$, \cite{AyonBeato:2009nh}. The rotating version of the AdS black hole solution can easily be obtained by
operating an improper boost in the $(t-\varphi)-$plane where $t$ (resp. $\varphi$) stands for
the time (resp. the angular) coordinate, and the resulting geometry turns out to be a rotating asymptotically
AdS solution of new massive gravity \cite{Oliva:2009ip}.  In the present work, we show that this  rotating solution of new massive gravity can support a conformal static stealth configuration given by a
nonminimally and conformally coupled scalar field. By conformal stealth configuration, we mean that the
solution only exists for $\xi=1/8$ and for a potential $U\propto \Phi^6$, which are the two conditions that ensure that the matter action
(\ref{sfnc}) enjoys the conformal symmetry.  We will clearly establish that the existence of
this stealth configuration is rendered possible because of the presence of a gravitational
hair in the black hole metric. Interestingly enough, we will also show that in the extremal case, the stealth scalar field diverges
at the horizon as it occurs for the conformal scalar field of the BBMB (Bocharova-Bronnikov-Melnikov-Bekenstein) solution in four dimensions, \cite{Bekenstein:1974sf,Bocharova:1970}.

The plan of this letter is as follows. In the next
section, we will give the stealth field equations (\ref{StealthEinsteineqs})  where the
gravity side is described by the equations of new massive gravity and, where the matter source part is given by the variation of
(\ref{sfnc}). We also present the rotating asymptotically AdS solution of new massive gravity as obtained in
Ref. \cite{Oliva:2009ip}. Then, in order to gain in clarity we first derive the
stealth configuration in the nonrotating case and present the extremal  solution. In the last part
of the section, we will show that the derivation of the rotating stealth solution is quite similar to the one
operated in the nonrotating case. Then, we will present the rotating asymptotically AdS black hole stealth configuration as well
as its extremal version. Finally, the last section will be devoted to the conclusions and future works.

%%%%%%%%%%%%%%%%%%%%%%%%%%%%%%%%%%%%%%%%%%%%%%%%%%%%%
\section{Field equations and stealth configurations}
%%%%%%%%%%%%%%%%%%%%%%%%%%%%%%%%%%%%%%%%%%%%%%%%%%%%%%%%%%%%
As said in the introduction, we consider as a gravity action the so-called new massive
gravity action \cite{Bergshoeff:2009hq} given by
\begin{equation}
S_G=\frac{1}{16\pi G}\int d^{3}x\sqrt{-g}\left[ R-2\lambda
-\frac{1}{m^{2}} \left( R_{\mu \nu }R^{\mu \nu
}-\frac{3}{8}R^{2}\right) \right],\label{eq:S}
\end{equation}
and, whose associated field equations read
\begin{equation}
R_{\mu \nu }-\frac{1}{2} R g_{\mu \nu }+\lambda {g}_{\mu \nu
}-\frac{1}{2m^{2}}K_{\mu \nu }=0,  \label{eq:NMG}
\end{equation}
where we have defined
\begin{eqnarray*}
K_{\mu \nu } &=&2\square {R}_{\mu\nu}-\frac{1}{2}\nabla_{\mu}
\nabla_{\nu }{R}-\frac{1}{2}\square {R}g_{\mu\nu}
+4R_{\mu\alpha\nu\beta}R^{\alpha\beta} \nonumber\\
&& -\frac{3}{2}RR_{\mu\nu}-R_{\alpha\beta}R^{\alpha\beta}g_{\mu\nu}
+\frac{3}{8}R^{2}g_{\mu \nu }.  \label{eom}
\end{eqnarray*}
It is easy to see that the field equations (\ref{eq:NMG}) admits solutions of constant curvature, that is
$R_{\alpha\beta}^{\mu\nu}=\Lambda \delta_{\alpha\beta}^{\mu\nu}$, with two different radii
\begin{eqnarray}
\Lambda_{\pm}=2m\left(m\pm\sqrt{m^2-\lambda}\right),
\end{eqnarray}
and these two radii coincide for
\begin{eqnarray}
m^2=\lambda:=-\frac{1}{2\,l^2}.
\label{sp}
\end{eqnarray}
At the special point (\ref{sp}) where the theory admits a unique maximally symmetric solution,
there exists a rotating asymptotically AdS  black hole solution which is locally
conformally flat \cite{Oliva:2009ip}. The metric is given by
\begin{equation}
ds^{2}=-N(r)F(r)dt^{2}+\frac{dr^{2}}{F(r)}+r^{2}\left(  d\varphi+N^{\varphi}(r)dt\right)
^{2}\ , \label{RBHM}
\end{equation}
where the structural metric functions $N$, $N^{\varphi}$ and $F$ read
\begin{eqnarray*}
&N(r)=\left[  1+\frac{bl^{2}}{4H}\alpha  \right]
^{2}\ ,\quad N^{\varphi}(r)=-\frac{a}{2r^{2}}\left(  4GM-bH\right)\nonumber\\
\label{Ns&F}\\
&F(r)=\frac{H^{2}}{r^{2}}\Big[  \frac{H^{2}}{l^{2}}+\frac{b}{2}\left(
2-\alpha\right)  H+\frac{b^{2}l^{2}}{16}\alpha^{2}-4GM (1-\alpha)\Big]  \ ,\nonumber
\end{eqnarray*}
and, where the function $H$ is defined by
\begin{eqnarray*}
H(r)=\left[  r^{2}-2GMl^{2}\alpha  -\frac{b^{2}l^{4}%
}{16}\alpha^{2}\right]^{\frac{1}{2}}\ .
\label{H}%
\end{eqnarray*}
In these expressions, the constant $\alpha$ is related to the rotation parameter $a$ through
\begin{eqnarray}
\alpha=1-(1-a^{2}/l^{2})^{1/2}.
\label{alpha}
\end{eqnarray}
Hence, this solution is described
by two constants related to the mass $M$ and the angular momentum $J=Ma$, while the constant $b$
which contributes to the expression of the mass can be viewed as a short of gravitational hair \cite{Giribet:2009qz,Perez:2011qp}.

We now investigate whether the black hole spacetime geometry (\ref{RBHM}) may accommodate a stealth configuration
given by a nonminimally coupled scalar field. To be more precise, we are interested on finding a static nontrivial scalar
field $\Phi=\Phi(r)$, and eventually a potential term such that the following equations
\begin{subequations}
\label{eqsmotion}
\begin{eqnarray}
&&G_{\mu \nu }+\lambda {g}_{\mu \nu
}-\frac{1}{2m^{2}}K_{\mu \nu }=0=T_{\mu\nu},\\
&&\Box \Phi = \xi R \Phi+\frac{d U}{d\Phi},
\end{eqnarray}
\end{subequations}
are satisfied on the rotating black hole background given by (\ref{RBHM}) at the special point (\ref{sp}). Here, $T_{\mu\nu}$ is the stress tensor associated to the variation of the matter
action (\ref{sfnc}), and is given by
\begin{eqnarray}
\label{tmunusf}
T_{\mu \nu}=&&\partial_{\mu}\Phi\partial_{\nu} \Phi-g_{\mu
\nu}\Big(\frac{1}{2}\,\partial_{\sigma}\Phi\partial^{\sigma}\Phi+U(\Phi)\Big)\nonumber\\
&&+\xi\left(g_{\mu \nu}\Box-\nabla_{\mu}\nabla_{\nu}+G_{\mu\nu}\right)\Phi^{2}.
\end{eqnarray}
We already know that the gravity part of the stealth equations (\ref{eqsmotion}) are satisfied on the
background (\ref{RBHM}) at the point (\ref{sp}), and so it only remains to solve the equations $T_{\mu\nu}=0$.

In order to gain in clarity, we first present the details of the computations in the
nonrotating case $a=0$ (or equivalently $\alpha=0$) and then we switch to  the rotating case. For a vanishing rotation parameter $a=0$,
the following combination
$T_t^t-T_{\varphi}^{\varphi}=0$ yields to a first-order
differential equation for the scalar field whose solution is given by
$$
\Phi_{\pm}(r)=\frac{C}{\sqrt{\pm(-br+8GM)}},
$$
where $C$ is an integration constant. Injecting this expression into the combination $T_r^r-T_{\varphi}^{\varphi}=0$, one obtains
the following constraint
$$
\frac{1}{4}\frac{\left(-r^2-brl^2+4GMl^2\right)}{(-br+8GM)^3l^2}\,C^2\,b^2\,(8\xi-1)=0,
$$
which is solved for $C=0$, or $b=0$, or $\xi=1/8$. However, the first two options implies that the scalar
field becomes constant and, hence in order to satisfy the constraint with a nontrivial scalar field, the nonminimal
coupling parameter must take its three-dimensional conformal value $\xi=1/8$. Finally, the remaining independent
equation given by the combination $T_t^t-T_r^r-T_{\varphi}^{\varphi}=0$ allows to express the potential term $U$ as a
local expression of the scalar field as
\begin{eqnarray}
U(\Phi)=\beta\,\Phi^6.
\label{potconf}
\end{eqnarray}
It is interesting to note that this  form of the potential together with the coupling $\xi=1/8$ are precisely those
that ensure the matter action (\ref{sfnc}) to be conformally
invariant. We then conclude that  the static solutions of the stealth
equations (\ref{eqsmotion}) on the nonrotating background (\ref{RBHM}) with $\alpha=0$  requires a conformal scalar
field source, and are given by
\begin{eqnarray}
\label{nonrotsol}
&&ds^2=-F(r)dt^2+\frac{dr^2}{F(r)}+r^2d\varphi^2,\nonumber\\
&& F(r)=\frac{r^2}{l^2}+br-4GM,\\
&&\Phi_{\pm}(r)=\left[\frac{16G^2M}{2\beta l^2}\left(M+\frac{b^2l^2}{16G}\right)\right]^{\frac{1}{4}}\frac{1}{\sqrt{\pm(-br+8GM)}}.\nonumber
\end{eqnarray}
It is evident that the existence of this stealth configuration is indicative to the presence
of the gravitational hair $b$ since in the limit $b=0$, the scalar field becomes constant. This result is not in
contradiction with the one of Ref. \cite{AyonBeato:2004ig} where the authors showed that in the BTZ case, $b=0$, the scalar
field must be nonstatic. In fact, the results obtained in Ref. \cite{AyonBeato:2004ig} together with the solution (\ref{nonrotsol}) clearly
establish the correlation between the presence of the gravitational hair $b$ and the possibility of having a
 nontrivial static stealth scalar field.

In the extremal case, that is for $M=-b^2l^2/(16G)$, the derivation of the stealth solution along the same line as before
implies again that $\xi=1/8$ but the potential
must be zero. The resulting extremal stealth solution reads
\begin{subequations}
\label{nonrotsolextr}
\begin{eqnarray}
&&ds^2=-\left(\frac{r}{l}+\frac{bl}{2}\right)^2dt^2+\frac{dr^2}{\left(\frac{r}{l}+\frac{bl}{2}\right)^2}+r^2d\varphi^2,\\
\nonumber\\
&&\Phi_{\pm}(r)=\frac{A}{\sqrt{\pm(2r+bl^2)}},
\end{eqnarray}
\end{subequations}
where now the scalar field depends on an arbitrary constant $A$. Various comments can be made concerning
this extremal solution.  Firstly, this latter can be obtained from the non extremal one (\ref{nonrotsol})
by taking the limit $M\to -b^2l^2/(16G)$ but at the same time $\beta\to 0$ such that $(M+b^2l^2/(16G))={\cal O}(\beta)$. In doing so, the
scalar field depends on an arbitrary constant. The occurrence of this arbitrary constant  can be easily explained since, in the absence of the potential
term, the energy-momentum tensor has a scaling symmetry, $\Phi\to \Omega\,\Phi$, where $\Omega$ is an arbitrary constant. Hence, the presence
of the arbitrary constant $A$ is just a consequence of this symmetry. We also stress that the scalar field
diverges at the horizon $r_+=-\frac{bl^2}{2}$ as it occurs for the BBMB solution in four dimensions.  This is intriguing in the sense
that the BBMB solution shares some features with this extremal stealth configuration. Indeed, the BBMB solution is a solution of a static
conformal scalar field in four dimensions without potential, and whose metric is also extremal (the extremal Reissner-Nordstrom spacetime).
The divergence of the BBMB scalar field at the horizon makes its physical
interpretation and the problem of its stability a subject of debate \cite{Bronnikov:1978mx,McFadden:2004ni}. A way of circumventing this problem
is to introduce a cosmological constant, whose effect is to precisely push this singularity behind the horizon,
as it has been done in Ref. \cite{Martinez:2002ru}. Because of these two examples (the BBMB solution and the extremal stealth configuration),
one is tempted to associate this pathology  to the extremal character of the metric together
with the conformal symmetry of the source.

Let us now consider the rotating case, $a\not=0$, for which the different steps to obtain  the stealth solution turn on to be analogous to those operated in the nonrotating case. Indeed, as before, the combination $T_t^t-T_{\varphi}^{\varphi}=0$ permits to obtain the expression of the scalar field while injecting this form into the combination $T_r^r-T_{\varphi}^{\varphi}=0$ yields to a rather complicated constraint. This latter is satisfied and yields to a nontrivial solution only in the case $\xi=1/8$. Finally, the combination $T_t^t-T_r^r-T_{\varphi}^{\varphi}=0$ allows to express the potential term $U$, and one obtains again the conformal potential (\ref{potconf}) while the remaining independent Einstein equations $T_{\varphi}^t=0$, being proportional to the combination $T_t^t-T_{\varphi}^{\varphi}=0$ is also satisfied. We end with the following rotating asymptotically anti de Sitter stealth configuration given
by the metric (\ref{RBHM}) together with the  scalar field
\begin{widetext}
\begin{eqnarray}
\Phi_{\pm}(r)=\left(\frac{256M^2G^2+16MGb^2l^2(\alpha+1)+b^4l^4\alpha}{2\beta l^2}\right)^{\frac{1}{4}}
\frac{1}{\sqrt{\pm\left(-b\sqrt{16r^2-32\alpha MGl^2-b^2l^4\alpha^2}+b^2l^2\alpha+32MG\right)}},
\end{eqnarray}
\end{widetext}
where $\alpha$ is defined in (\ref{alpha}). It is evident that in the vanishing rotation limit $a\to 0$
(or equivalently $\alpha\to 0$), this solution reduces to (\ref{nonrotsol}). Moreover, as in the nonrotating case, we emphasize again
that this is
the presence of the gravitational hair $b$ that prevents the scalar field to be trivial. The
extremal version for $M=-b^2l^2/(16G)$  requires the nonminimal coupling $\xi$ to be
the conformal one $\xi=1/8$ as well as a vanishing potential $U=0$, and the scalar field is given by
\begin{eqnarray*}
\Phi(r)=\frac{A}{\sqrt{\pm\left(-b\sqrt{16r^2+b^2l^4\alpha(2-\alpha)}+l^2b^2(\alpha-2)\right)}},
\end{eqnarray*}
where $A$ is an arbitrary constant. As in the nonrotating case, the scalar field diverges at the horizon and
this configuration can be obtained by taking the limits $M\to -b^2l^2/(16G)$ and $\beta\to 0$ such that $(M+b^2l^2/(16G))={\cal O}(\beta)$.

%%%%%%%%%%%%%%%%%%%%%%%%%%%%%%%%%%%%%
\section{Comments and conclusions}
%%%%%%%%%%%%%%%%%%%%%%%%%%%%%%%%%%%%%
Here, we have shown that the rotating asymptotically anti de Sitter black hole solution of new massive gravity
in three dimensions can support a nontrivial static stealth configuration given by a nonminimally and conformally
coupled scalar field. We have clearly established that this is the presence of the gravitational hair $b$
that prevents the scalar field to be trivial. The extremal version of this stealth configuration presents the
same pathology (namely the divergence of the scalar field at the horizon) that the BBMB solution in four dimensions.

There are many issues related to the present work that will be interesting to explore but we would like to emphasize the
thermodynamics issue. Indeed, since we have obtained a
black hole solution, it is natural to wonder about the thermodynamics. However, in order to compute the mass, the temperature and the entropy of this solution, we are faced with the following problem.
In fact, one may note that the stealth
equations (\ref{eqsmotion}) can be viewed as a particular solution of the field equations
arising from the variation of the action
\begin{eqnarray}
S=S_G+ S_M.
\label{fullaction}
\end{eqnarray}
where $S_G$ is the new massive gravity action (\ref{eq:S}) and $S_M$ is the source action (\ref{sfnc}). For simplicity, let us consider the
nonrotating case. The temperature is
given as in the free source case \cite{Giribet:2009qz,Perez:2011qp} by
\begin{eqnarray}
T=\frac{1}{\pi l}\sqrt{GM+\frac{b^2l^2}{16}}
\end{eqnarray}
while the entropy ${\cal S}$ computed with the help of the Wald formula \cite{Wald:1993nt} yields
\begin{eqnarray}
{\cal S}
=\frac{2\pi l}{\sqrt{G}}\sqrt{M+\frac{b^2l^2}{16 G}}-
\frac{\pi}{8}\sqrt{\frac{2GM}{\beta}}.
\end{eqnarray}
However, a simple computation shows that the product $Td{\cal S}$ 
is not a total derivative, and hence we are faced with the problem that the first law is not 
satisfied  unless there is some additional charge to be considered. It will be interesting to further explore the 
thermodynamic issue of the stealth solutions found here.

Other questions can be asked related to this work as for example what is the precise role of the
gravitational hair in the emergence of such configuration. Also, we have been interested on looking
only for static stealth configuration that can be supported by the
rotating solution of new massive gravity. From Ref. \cite{AyonBeato:2004ig}, we learn that in the BTZ case, the stealth
scalar field configuration must be nonstatic and the angular momentum must be zero. We may ask whether there exists a nonstatic
 stealth configuration in the case of the rotating solution of new massive gravity. Finally, it is also natural to explore if
 our results can be extended in higher dimensions.

\begin{acknowledgments}
We thank Moises Bravo, Julio Oliva, Tahsin Sisman, David Tempo and Ricardo Troncoso  for useful discussions. MH is partially
supported by grant 1130423 from FONDECYT, by grant ACT 56 from
CONICYT and from CONICYT, Departamento de Relaciones Internacionales
``Programa Regional MATHAMSUD 13 MATH-05''.
\end{acknowledgments}

%%%%%%%%%%%%%%%%%%%%%%%%%%%


\begin{thebibliography}{99}
%%%%%%%%%%%%%%%%%%%%%%%%%%%



\bibitem{AyonBeato:2004ig}
  E.~Ayon-Beato, C.~Martinez and J.~Zanelli,
  ``Stealth scalar field overflying a (2+1) black hole,''
  Gen.\ Rel.\ Grav.\  {\bf 38}, 145 (2006)
  [hep-th/0403228].
  %%CITATION = HEP-TH/0403228;%%
  %19 citations counted in INSPIRE as of 14 Nov 2013


\bibitem{Banados:1992wn}
  M.~Banados, C.~Teitelboim and J.~Zanelli,
  ``The Black hole in three-dimensional space-time,''
  Phys.\ Rev.\ Lett.\  {\bf 69}, 1849 (1992)
  [hep-th/9204099].
  %%CITATION = HEP-TH/9204099;%%
  %1558 citations counted in INSPIRE as of 06 Aug 2013





\bibitem{AyonBeato:2005tu}
  E.~Ayon-Beato, C.~Martinez, R.~Troncoso and J.~Zanelli,
  ``Gravitational Cheshire effect: Nonminimally coupled scalar fields may not curve spacetime,''
  Phys.\ Rev.\ D {\bf 71}, 104037 (2005)
  [hep-th/0505086].
  %%CITATION = HEP-TH/0505086;%%
  %26 citations counted in INSPIRE as of 14 Nov 2013


%\cite{Gaete:2013ixa}
\bibitem{Gaete:2013ixa}
  M.~B.~Gaete and M.~Hassaine,
  ``Topological black holes for Einstein-Gauss-Bonnet gravity with a nonminimal scalar field,''
  arXiv:1308.3076 [hep-th].
  %%CITATION = ARXIV:1308.3076;%%







%\cite{Gaete:2013oda}
\bibitem{Gaete:2013oda}
  M.~B.~Gaete and M.~Hassaine,
  ``Planar AdS black holes in Lovelock gravity with a nonminimal scalar field,''
  arXiv:1309.3338 [hep-th].
  %%CITATION = ARXIV:1309.3338;%%



 \bibitem{Bergshoeff:2009hq}
  E.~A.~Bergshoeff, O.~Hohm and P.~K.~Townsend,
  ``Massive Gravity in Three Dimensions,''
  Phys.\ Rev.\ Lett.\  {\bf 102}, 201301 (2009)
  [arXiv:0901.1766 [hep-th]].
  %%CITATION = ARXIV:0901.1766;%%
  %304 citations counted in INSPIRE as of 14 Nov 2013


\bibitem{Oliva:2009ip}
  J.~Oliva, D.~Tempo and R.~Troncoso,
  ``Three-dimensional black holes, gravitational solitons, kinks and wormholes for BHT massive gravity,''
  JHEP {\bf 0907}, 011 (2009)
  [arXiv:0905.1545 [hep-th]].
  %%CITATION = ARXIV:0905.1545;%%
  %65 citations counted in INSPIRE as of 08 Nov 2013


\bibitem{AyonBeato:2009nh}
  E.~Ayon-Beato, A.~Garbarz, G.~Giribet and M.~Hassaine,
  ``Lifshitz Black Hole in Three Dimensions,''
  Phys.\ Rev.\ D {\bf 80}, 104029 (2009)
  [arXiv:0909.1347 [hep-th]].
  %%CITATION = ARXIV:0909.1347;%%
  %75 citations counted in INSPIRE as of 14 Nov 2013


\bibitem{Bekenstein:1974sf}
  J.~D.~Bekenstein,
  ``Exact Solutions Of Einstein Conformal Scalar Equations,''
  Annals Phys.\  {\bf 82}, 535 (1974).
%%CITATION = APNYA,82,535;%%

\bibitem{Bocharova:1970}
N.~M.~Bocharova, K.~A.~Bronnikov and V.~N.~Melnikov, ``An exact
solution of the system of Einstein equations and mass-free scalar
field,'' Vestn. Mosk. Univ. Fiz. Astron. {\bf 6} (1970) 706 [Moscow
Univ. Phys. Bull. {\bf 25} (1970) 80].

\bibitem{Giribet:2009qz}
  G.~Giribet, J.~Oliva, D.~Tempo and R.~Troncoso,
  ``Microscopic entropy of the three-dimensional rotating black hole of BHT massive gravity,''
  Phys.\ Rev.\ D {\bf 80}, 124046 (2009)
  [arXiv:0909.2564 [hep-th]].
  %%CITATION = ARXIV:0909.2564;%%
  %25 citations counted in INSPIRE as of 14 Nov 2013

\bibitem{Perez:2011qp}
  A.~Perez, D.~Tempo and R.~Troncoso,
  %``Gravitational solitons, hairy black holes and phase transitions in BHT massive gravity,''
  JHEP {\bf 1107}, 093 (2011)
  [arXiv:1106.4849 [hep-th]].
  %%CITATION = ARXIV:1106.4849;%%
  %11 citations counted in INSPIRE as of 14 Nov 2013


\bibitem{Bronnikov:1978mx}
  K.~A.~Bronnikov and Y.~.N.~Kireev,
  ``Instability of Black Holes with Scalar Charge,''
  Phys.\ Lett.\ A {\bf 67} (1978) 95.
  %%CITATION = PHLTA,A67,95;%%

%\cite{McFadden:2004ni}
\bibitem{McFadden:2004ni}
  P.~L.~McFadden and N.~G.~Turok,
  ``Effective theory approach to brane world black holes,''
  Phys.\ Rev.\ D {\bf 71} (2005) 086004
  [hep-th/0412109].
  %%CITATION = HEP-TH/0412109;%%

\bibitem{Martinez:2002ru}
  C.~Martinez, R.~Troncoso and J.~Zanelli,
  ``De Sitter black hole with a conformally coupled scalar field in  four
  dimensions,''
  Phys.\ Rev.\  D {\bf 67}, 024008 (2003)
  [arXiv:hep-th/0205319].
  %%CITATION = PHRVA,D67,024008;%%




\bibitem{Wald:1993nt}
  R.~M.~Wald,
  ``Black hole entropy is the Noether charge,''
  Phys.\ Rev.\ D {\bf 48}, 3427 (1993)
  [gr-qc/9307038].
  %%CITATION = GR-QC/9307038;%%
  %823 citations counted in INSPIRE as of 14 Nov 2013


\end{thebibliography}
\end{document}